\documentclass[11pt]{article}
\usepackage{graphicx}
\usepackage{dcolumn}   
\usepackage{bm}        
\usepackage{amssymb}   
\usepackage{pslatex}

\title{Jamming phase diagram of athermal emulsions with short-range attraction}

\author{Ivane Jorjadze, Lea-Laetitia Pontani, Katherine A. Newhall, Jasna Bruji\'c\\
\small Center for Soft Matter Research, Physics Department, New York University\\[-0.8ex]
\small New York University, 4 Washington Place, New York, NY 10003\\
}

\begin{document}
\maketitle

\begin{abstract}
Using confocal microscopy we investigate the effect of attraction on the packing of polydisperse emulsions under gravity. We find that the distributions of neighbors, coordination number, and local packing fraction as a function of attraction are captured by recently proposed geometrical modeling and statistical mechanics approaches to granular matter. This extends the range of applicability of these tools to polydisperse, attractive jammed packings. Furthermore, the dependence of packing density and average coordination number on the strength of attraction provides the first experimental test of a phase diagram of athermal jammed particles. The success of these theoretical frameworks in describing a new class of systems gives support to the much-debated statistical physics of jammed matter.
\end{abstract}

The random packing of particles is a problem of great practical and theoretical importance with applications to systems as diverse as granular materials, emulsions and colloids. How athermal particles pack is affected by many parameters, including the packing protocol, the shape and roughness of the particles, as well as their polydispersity. Nevertheless, experiments and numerical simulations have suggested the existence of reproducible disordered packing structures, which can be achieved by sedimenting and shaking monodisperse \cite{scott1960,finney1970} and polydisperse \cite{clus2009} spheres, ellipsoids \cite{chaikEl2005} and tetrahedra \cite{jaosh2010}. For frictional particles, there emerges a reversible packing density curve from random loose packing (RLP) to random close packing (RCP) as a function of the applied tapping protocol \cite{nowak1997}. The dependence of density of these jammed states on the friction coefficient has led to an equation of state and a prediction of a jamming phase diagram to describe the behavior of these systems in numerical simulations \cite{makse2008}. The reproducibility of all these jammed states for a given set of experimental conditions suggests that a statistical mechanics framework may apply to these inherently out of equilibrium systems \cite{edwards1989}. Other studies have argued that the history dependence and heterogeneities in the packing preclude a general theory of jammed granular matter\cite{torq2000,ohern2001}.

In this paper, we introduce a novel route to jamming by packing athermal, frictionless spheres under gravity in the presence of a short-range attractive force.
By examining the packing geometry in 3D we find that the strength of the attractive interaction controls the packing density and the microstructure of jammed states. We thus provide the first experimental verification of another branch of the phase diagram of jammed matter originally predicted for monodisperse repulsive particles \cite{makse2008}, namely that of polydisperse attractive particles. Moreover, our experimental distributions of local packing properties are in very good agreement with the predictions of recently proposed models applied to repulsive particle aggregates: the `granocentric' model \cite{clus2009} and the `$k$-gamma distribution' of the local packing fraction \cite{aste2008}. This study therefore extends the limits of applicability of geometrical modeling and statistical mechanics approaches to capture the effects of both attraction and polydispersity. These results open the way to use tools originally developed to investigate granular materials to a new class of systems.

Jamming attractive athermal particles under gravity should not be confused with the widely studied gel transition in thermal systems, where classical thermodynamics applies \cite{bib1992,weitz2008,poon2004}. In thermal systems attractive forces lead to colloidal aggregates with fractal structures at arbitrarily low densities, while here the absence of thermal fluctuations seems to localize the effect of attraction to the first shell of neighbors where the forces due to gravity and attraction balance to form a mechanically stable pack. Despite the fundamental differences in the mechanism of their creation and the resulting structures, this study reveals one important similarity. We observe a reentrant behavior of packing density with attraction, analogous to that observed in thermal glassy systems \cite{poon2004, sperl2004}, thus questioning its postulated dynamical origin.


To create a jammed packing we cream athermal oil-in-water emulsions under gravity, as described in reference \cite{jasna2003}. In addition, we introduce an attractive potential by varying the concentration of sodium dodecyl sulfate (SDS) micelles, which act as a depletant \cite{bib1999}. The depletion energy has an entropic origin and its dependence on the surfactant concentration is derived in reference \cite{asac1958} and validated for our emulsion system in \cite{bib1999, bib1990, fas2005, yod1999}. We therefore calculate the short-range attractive force $F_{\textrm{d}}$ at an equilibrium distance of around $\textrm{10 nm}$ \cite{bib1999} and using a critical micellar concentration (CMC) of $\textrm{13 mM}$ \cite{jasna2007}. The range of attraction is limited by the saturation point of SDS (i.e. $\textrm{50 mM}$) at a maximum average force of $\langle F_{\textrm{d}}\rangle=33 \textrm{ pN}$. This average depletion force is defined for each packing as the force between two droplets of average radius, $\langle r\rangle= 3.5 \textrm{ } \mu \textrm{m}$. Since the particle size distribution has a 25\% standard deviation in radius, the depletion force between particle pairs picked at random, calculated using the standard AO theory \cite{asac1958}, has a distribution with $\pm{20\%}$ standard deviation. This calculation is justified because the the micelles are approximately $500$ times smaller than the size of the smallest emulsion droplet and the micelle volume fraction is always below $2\%$~\cite{fas2005, yod1999}. The average attractive forces $\langle F_{\textrm{d}}\rangle$ studied here range from $10$ to $100$ times stronger than the weight of a single particle. This corresponds to small deformations in the range of $1-10 \textrm{ \AA}$, such that the spherical approximation holds.

\begin{figure}
\begin{center}
\includegraphics[scale=0.2]{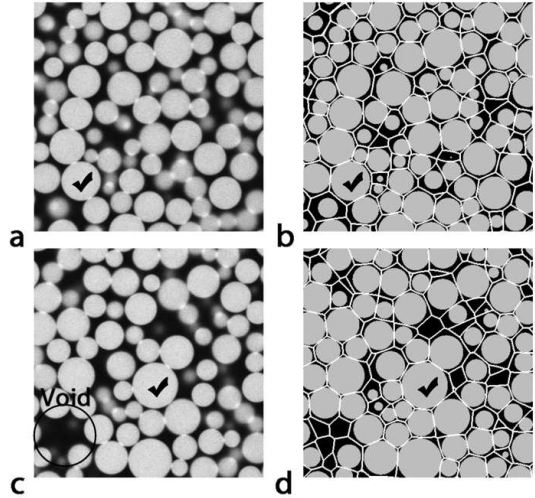}
\end{center}
\caption{\label{fig1} Confocal 2D slices and their reconstructed images for repulsive (a, b) and attractive (c, d) 3D emulsions. Ticks mark the original and reconstructed spheres. Navigation maps (white lines) are superimposed with the packing (b, d). Voids shown in (c) appear to occupy cells as particles above and below the confocal plane share void space in the third dimension.}
\end{figure}

The protocol to prepare the jammed packing involves creaming a very dilute emulsion at a density $\phi\approx5\%$ to avoid aggregation of the particles. Refractive index matching the droplets and the aqueous phase allows us to image the dynamics of the packing process 
using a fast scanning confocal microscope (Leica TCS SP5 II).
The voxel size is $\textrm{130 nm}$ in the horizontal $xy$ plane and $\textrm{300 nm}$ in the vertical $z$ direction. We image a box size of $65\times65\times100\mu m$, several particles away from the boundaries of a $1mL$ container. Once the packing of $\approx1500$ droplets is formed, confocal images in Figs.~\ref{fig1}a, c reveal a visual difference between the attractive and repulsive jammed packings: repulsive packings display homogeneous structures while high enough attraction gives rise to voids within the packing. We analyse the structure in terms of the particle positions and radii with subvoxel accuracy using a Fourier transform algorithm \cite{jasna2003}. We then use the geometrical overlaps in the reconstructed spheres (see Figs.~\ref{fig1}b, d) to identify particles that are in contact. This allows us to measure the number of contacts a particle has, i.e. its coordination number, $z$. Error in estimating particle positions and radii translates to an error of $\pm 0.3$ in the $\langle z \rangle$ estimation.
To characterize the local neighborhood of each particle we tessellate space using the navigation map \cite{medvedev1995,clus2009} shown in Figs.~\ref{fig1}b, d. This mapping defines two local quantities: the number of neighbors, $n$, a particle shares an interface with and the local packing fraction, $\phi_{\textrm{loc}}$, which is the ratio between the particle volume, $V_{\textrm{p}}$, and the volume of the corresponding cell, $V$. The global density, $\phi$, is determined by the total volume of the particles divided by the system volume.

\begin{figure*}
\begin{center}
\includegraphics[scale=0.45]{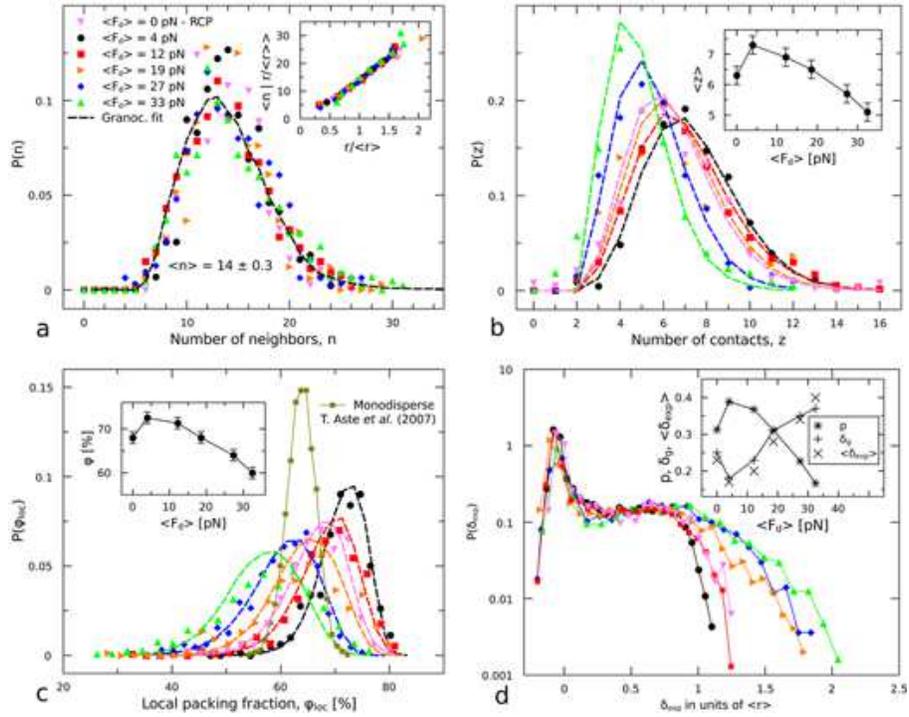}
\end{center}
\caption{\label{fig2} (Color online) Probability density distributions $P(n)$ in (a), $P(z)$ in (b), and $P(\phi_{\textrm{loc}})$ in (c) are shown as a function of $\langle F_{\textrm{d}}\rangle$, while the insets in (a, b) present the dependency of the average values. The legend in (a) defines the colors and symbols for all data sets, which are used throughout the manuscript. Repulsive emulsion packings at RCP are included for comparison. All distributions are fitted by the granocentric model, whose parameters are shown in the inset in (c). The global trends in $\langle F_{\textrm{d}}\rangle$ and $\langle z \rangle$ as a function of $\phi$ are shown in (d).}
\end{figure*}

In Fig.~\ref{fig2} we present the probability density distributions of the local parameters $z$, $n$, and $\phi_{\textrm{loc}}$ as a function of attraction. Even though the packings exhibit more void space as the attraction is increased, the neighbor number distribution $P(n)$ shown in Fig.~\ref{fig2}a, is independent of attraction. Just as in the repulsive packings, the $\langle n \rangle$ is $14 \pm 0.3$ with a standard deviation of $30\%$. This makes sense because the navigation map partitions the extra void space between the neighbors, but the number of common interfaces remains the same. The inset shows that the particle size dependence on the number of neighbors also remains the same for all attractions, rendering the number of neighbors an insensitive measure of the packing structure. 
The coordination number $z$ reports on the number of force-bearing particles in the mechanically stable packing, which is greatly affected by the attractive forces in addition to gravity. The measured distributions $P(z)$ in Fig.~\ref{fig2}b peak at decreasing values from $7$ to $4$ and become narrower as a function of attraction \cite{lee2006}. 
The isostatic condition predicts a value of $\langle z \rangle = 6$ for repulsive frictionless spheres in 3D \cite{alexander1998}, yet the inset in Fig.~\ref{fig2}b shows that $\langle z \rangle$ decays as a function of $\langle F_{\textrm{d}}\rangle$ to below this value. 
With increasing attraction the probability of such local configurations also increases, as shown in the $P(z)$, thus shifting $\langle z \rangle$. Analogously, the presence of stable loose voids shifts the distributions of local packing fraction $P(\phi_{\textrm{loc}})$ shown in Fig.~\ref{fig2}c towards lower packing densities with wider distributions as a function of increasing attraction. The decrease in global density $\phi$ is also apparent in the top panel of Fig.~\ref{fig2}d.

On the other hand, the initial increase in $\langle z \rangle$ above isostaticity and $\phi$ above RCP at the onset of attraction, shown in the graphs to be significantly larger than the error bars on the measurements, remains puzzling. While jammed random states above RCP have been proposed numerically in athermal systems \cite{torq2000}, this is the first experimental system to explore such compact configurations that avoid crystallization. Whether these denser states become accessible due to polydispersity remains an open question. This turnover implies that isostatic packings with a density at RCP can also be achieved with a nonzero attractive potential, as shown by the magenta points and those at $19 \textrm{ pN}$ in both panels in Fig.~\ref{fig2}d. Interestingly, a similar turnover in $F_{\textrm{d}}$ versus $\phi$ is seen in attractive colloidal gels undergoing structural arrest and interpreted as the signature of a reentrant glass transition \cite{poon2004, sperl2004}.

To understand the observed trends in the probability distributions in Fig.~\ref{fig2} quantitatively, we interpret the data using the granocentric local model introduced in \cite{clus2009}. 
The model uses as inputs the experimental values of $\langle n \rangle$, $\langle z \rangle$ (excluding particles with $z\le3$) and global density $\phi$, and outputs the distance $\delta_{\textrm{g}}$ between two neighboring sphere surfaces, as well as the distributions $P(n)$, $P(z)$, and $P(\phi_{\textrm{loc}})$, shown in Fig.~\ref{fig2}a,b,c. In the inset in Fig.~\ref{fig2}c, the ratio of contacts to neighbors, given by $p=(\langle z \rangle-3)/(\langle n \rangle-3)$ is shown to decrease as a function of the attractive force due to the decrease in $\langle z \rangle$, while the predicted distance $\delta_{\textrm{g}}$ naturally increases as the packings become looser and agrees very well with the average experimental value $\langle \delta_{\textrm{exp}} \rangle$. Further confidence in the model is shown in Figs.~\ref{fig2}a, b, c by the agreement between the measured distributions and the theoretical predictions in which all parameters are fixed by experimentally determined  values.
It is surprising that a local stochastic model captures all the probability distributions, since it assumes that there are no local correlations between neighboring particles nor long range correlations beyond the first shell of neighbors. This result is in contrast to the structure of attractive colloidal gels, where the fractal dimension indicates correlations between thermal particles that persist throughout the system \cite{bib1992, weitz2008}.

\begin{figure*}
\begin{center}  
\includegraphics[scale=0.45]{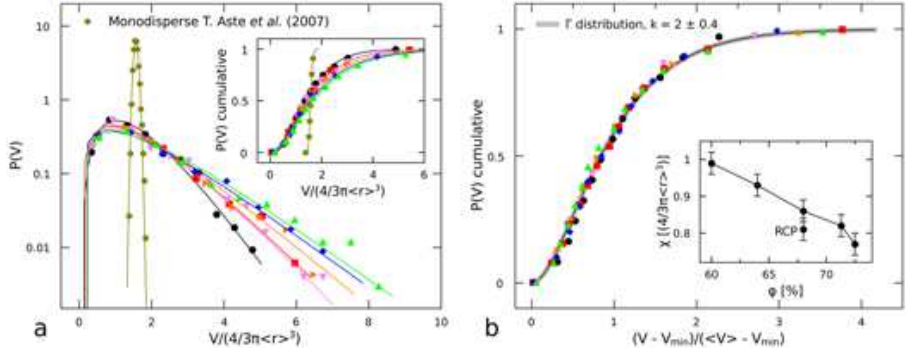}
\caption{\label{fig3} (Color online) (a) Evolution of $P(V)$ and its cumulative distribution for different levels of attraction. Notice that $V$ is rescaled by the volume of the average particle for each distribution. (b) Data collapse is observed when $V$ is rescaled by the shifted average volume of the cell $\langle V\rangle$. The inset shows the compactivity $\chi$ decrease with increasing density $\phi$.}
\end{center}
\end{figure*}

Given the success of the local granocentric description of the packing structure, we next consider the fluctuations in the volume $V$ of each cell in the navigation map in terms of a statistical mechanics framework \cite{edwards1989}.
Independently picking volumes $V$ from a uniform distribution results in an exponential probability distribution of volumes $P(V)$ in the thermodynamic limit. This Boltzmann-type distribution maximizes the entropy given the constraints.
Assuming that each cell $V$ is made of $k$ elementary cells \cite{aste2007,aste2008}, this distribution becomes a shifted $k$-gamma distribution with a shape parameter~$k$ and a scale parameter $\langle V \rangle - V_{\textrm{min}}$:
\begin{equation} P(V) =\frac{k^k}{\Gamma(k)} \frac{\left( V - V_{\textrm{min}} \right)^{(k-1)}}{\left(\langle V\rangle - V_{\textrm{min}} \right)^k} \exp{\left( -k \frac{V - V_{\textrm{min}}}{\langle V\rangle - V_{\textrm{min}}} \right)}. \label{gamma} \end{equation}
Indeed, a wide range of monodisperse packings with different packing protocols and global densities have been successfully fit by this type of distribution \cite{aste2007,aste2008,weeks2010,dauch2006}. Here we investigate the effect of polydispersity and subsequently attraction on~$P(V)$. Fig.~\ref{fig3}a shows that polydispersity of $25\%$ alone significantly broadens the distribution compared to the previously measured monodisperse case (star symbols) \cite{aste2007} and shifts the global packing fraction to a higher value.
Rescaling each $V$ by the volume of the particle itself leads to a broader distribution $P(\phi_{\textrm{loc}})$ in the polydisperse case, shown in Fig.~\ref{fig2}c, indicative of an additional source of randomness to the size distribution.
Fitting the mono and polydisperse cases with $k\approx14$ and $k\approx2$, respectively, captures the observed changes in shape and width. This means that
the polydisperse packing approaches the Boltzmann distribution.
As a function of attraction, Fig.~\ref{fig3}a shows an increase in both $\langle V\rangle$ and the standard deviation $\sigma$ of the distributions, as expected from the looser packing structures. Since Eq.~(\ref{gamma}) gives $(\langle V \rangle-V_{\textrm{min}})/\sigma=\sqrt{k}$, this dependence leads to a constant value of $k = 2.0 \pm 0.4 $ for all levels of attraction. This is shown in Fig.~\ref{fig3}b by the data collapse that results from rescaling the volumes as $(V - V_{\textrm{min}})/(\langle V\rangle - V_{\textrm{min}})$.

Within the thermodynamic framework, a simple counting argument of volumes leads to a definition of the entropy $S$ and consequently an analogue of inverse temperature $\chi^{-1} = \partial{S} / \partial{V_{\textrm{tot}}}$, which is known as the compactivity in granular statistical mechanics. The resulting expression for $\chi=(\langle V \rangle-V_{\textrm{min}})/k$ is shown to be linearly decreasing with density in the inset in Fig.~\ref{fig3}b. 
Recently, a phase diagram for jammed matter have been proposed in terms of the dependence of $\chi$ and $\langle z \rangle$ on density for monodisperse frictional hard spheres \cite{makse2008}. In comparison, our data shown in Fig.~\ref{fig2}d lies below the predicted RLP line at infinite $\chi$ (dash-point line), which we have extended beyond monodisperse RCP for comparison. We achieve packings denser than RLP over the same range of $\langle z\rangle$, consistent with the fact that we explore states with decreasing values of $\chi$. This also shows that the line we observe is not simply the RLP line shifted due to polydispersity. Surprisingly, we extend the limits of the phase diagram to hyperstatic packings up to $ \langle z \rangle = 7.5 $ and densities up to $74\%$, which is above the RCP density ($68\%$ in the polydisperse case). The microscopic mechanism for jamming with attraction and the origin of these states provide interesting questions for further study.

\bibliography{mybib}
\end {document}